\documentclass[12pt,a4paper]{article}
\usepackage{amsmath}
\usepackage{latexsym}
\makeatletter
\def\rddots{\mathinner{\mkern1mu\raise\p@%
    \vbox{\kern7\p@\hbox{.}}\mkern2mu%
    \raise4\p@\hbox{.}\mkern2mu\raise7\p@\hbox{.}\mkern1mu}}
\makeatother
\newcommand{\ket}[1]{{\vert{#1}\rangle}}

\begin{document}

\title{\sl An Approximate Solution of the Dynamical Casimir Effect 
in a Cavity with a Two--Level Atom}
\author{
  Kazuyuki FUJII
  \thanks{E-mail address : fujii@yokohama-cu.ac.jp }\quad and\ \ 
  Tatsuo SUZUKI
  \thanks{E-mail address : suzukita@sic.shibaura-it.ac.jp }\\
  ${}^{*}$International College of Arts and Sciences\\
  Yokohama City University\\
  Yokohama, 236--0027\\
  Japan\\
  ${}^{\dagger}$Department of Mathematical Sciences\\
  College of Systems Engineering and Science\\
  Shibaura Institute of Technology\\
  Saitama, 337--8570\\
  Japan\\
  }
\date{}
\maketitle
\begin{abstract}
  In this paper we treat the so--called dynamical Casimir effect 
in a cavity with a two--level atom and give an analytic approximate 
solution under the general setting. 

The aim of the paper is to show another approach based on 
Mathematical Physics to the paper 
[arXiv : 1112.0523 (quant-ph)] by A. V. Dodonov and V. V. Dodonov. 
We believe that our method is simple and beautiful.
\end{abstract}

\vspace{5mm}\noindent

\section{Introduction}
In this paper we revisit the so-called dynamical Casimir effect (DCE). 
This means the photon generation from vacuum due to the motion 
(change) of neutral boundaries, which corresponds to a kind of 
quantum fluctuation of the electro-magnetic field. 

This phenomenon is a typical example of interactions between the microscopic 
and the macroscopic levels and is very fascinating   
from the point of view of not only (pure) Physics but also Mathematical Physics. 
See for example \cite{Dodonov} and its references.

Then, how do we detect the photons generated ? This is also an important 
problem. Recently, Dodonov and Dodonov in \cite{Dodonovs} treated 
this problem by use of a two-level system of an atom in a cavity. 
They called it ``the cavity dynamical Casimir effect in the presence of 
a two-level atom", and gave an effective Hamiltonian by simplifying 
the Hamiltonian given by Law \cite{Law} and adding a two-level system 
of an atom to it, and constructed an approximate analytical solution.

In this paper we treat this problem once more and present another 
approach to construct an analytic approximate solution 
under the general setting. 
We believe that our method is clearer than that of \cite{Dodonovs}.

\section{Model}
First of all let us make a brief review of \cite{Dodonovs} within 
our necessity. In the following we set $\hbar=1$ for simplicity.

\vspace{5mm}\noindent
(i)\ Cavity DCE (\cite{Law}, \cite{Dodonov}). \ For this we take 
the simplest Hamiltonian which is the special case of Law 
\cite{Law} (namely, $\epsilon(x,t)=\epsilon(t)$)
\begin{equation}
\label{eq:Law}
H_{DCE}=\omega(t)a^{\dagger}a+i\chi(t)\left\{(a^{\dagger})^{2}-a^{2}\right\}
\end{equation}
where $\omega(t)$ is a periodic function depending on the cavity form and 
$\chi(t)$ is given by
\[
\chi(t)
=\frac{1}{4\omega(t)}\frac{d\omega(t)}{dt}
=\frac{1}{4}\frac{d}{dt}\log|\omega(t)|
\]
, and $a$ and $a^{\dagger}$ are the cavity photon annihilation and 
creation operators respectively. 
Therefore, the physics that we are treating is two-photon generation 
processes from the vacuum state.

\vspace{5mm}\noindent
(ii)\ Detection (\cite{WS}, \cite{JC}). \ For this we take a two-level system 
of an atom inserted in the cavity and the Rabi Hamiltonian as interaction
\begin{eqnarray}
\label{two-level system}
H_{D}
&=&\frac{\Omega}{2}\sigma_{3}\otimes {\bf 1}+
g(\sigma_{+}+\sigma_{-})\otimes (a+a^{\dagger})   \nonumber \\
&=&\frac{\Omega}{2}\sigma_{3}\otimes {\bf 1}+
g(\sigma_{+}\otimes a+\sigma_{+}\otimes a^{\dagger}+
  \sigma_{-}\otimes a+\sigma_{-}\otimes a^{\dagger}) 
\end{eqnarray}
where ${\bf 1}$ is the identity operator on the Fock space 
generated by $\{a, a^{\dagger}, N\equiv a^{\dagger}a\}$ and
\[
\sigma_{3}=
\left(
\begin{array}{cc}
1 & 0   \\
0 & -1
\end{array}
\right),\quad
\sigma_{+}=
\left(
\begin{array}{cc}
0 & 1  \\
0 & 0 
\end{array}
\right),\quad
\sigma_{-}=
\left(
\begin{array}{cc}
0 & 0  \\
1 & 0 
\end{array}
\right),\quad
1_{2}=
\left(
\begin{array}{cc}
1 & 0  \\
0 & 1 
\end{array}
\right).
\]

However, to solve this system is very hard. In fact, our aim is to obtain 
some analytic approximate solution. Therefore, as a rule 
we replace the Rabi Hamiltonian with the Jaynes-Cummings one
\begin{equation}
\label{eq:reduction form}
\widetilde{H}_{D}
=\frac{\Omega}{2}\sigma_{3}\otimes {\bf 1}+
g(\sigma_{+}\otimes a+\sigma_{-}\otimes a^{\dagger}).
\end{equation}

Here, let us note the well--known $su(2)$ relations
\[
[(1/2)\sigma_{3},\sigma_{+}]=\sigma_{+},\quad
[(1/2)\sigma_{3},\sigma_{-}]=-\sigma_{-},\quad
[\sigma_{+},\sigma_{-}]=2\times (1/2)\sigma_{3}.
\]

\vspace{5mm}\noindent
(iii)\ Total System. \ By adding (\ref{eq:reduction form}) 
to (\ref{eq:Law}) we treat the (effective) Hamiltonian like
\begin{eqnarray}
\label{eq:hamiltonian-1}
H&=&H(t)=H_{DCE}+\widetilde{H}_{D} \nonumber \\
&=&
\omega(t)1_{2}\otimes N+i\chi(t)1_{2}\otimes \left\{(a^{\dagger})^{2}-a^{2}\right\}+
\frac{\Omega}{2}\sigma_{3}\otimes {\bf 1}+g(\sigma_{+}\otimes a+\sigma_{-}\otimes a^{\dagger}).
\end{eqnarray}

Our aim is to solve the Schr{\"o}dinger equation
\begin{equation}
\label{eq:Schrodinger}
i\frac{d}{dt}\ket{\Psi(t)}=H\ket{\Psi(t)}=H(t)\ket{\Psi(t)}
\end{equation}
under the general setting.

If $H$ is time--independent, then the general (formal) solution 
is given by
\[
\ket{\Psi(t)}=e^{-itH}\ket{\Psi(0)}.
\]
To calculate $e^{-itH}$ exactly is another problem 
(which is in general very hard).  
However, in our case $H$ is time--dependent, so 
solving (\ref{eq:Schrodinger}) becomes increasingly difficult. 
We must give a further approximation to the Hamiltonian.

\vspace{5mm}\noindent
(iv)\ Interaction Picture. \ From here let us change to 
the interaction picture. Namely, for $V=V(t)$ we set
\begin{equation}
\ket{\Phi(t)}=V^{\dagger}\ket{\Psi(t)}
\ \Longleftrightarrow\ \ket{\Psi(t)}=V\ket{\Phi(t)}.
\end{equation}
Then it is easy to see that the equation (\ref{eq:Schrodinger}) can 
be changed to
\begin{equation}
\label{eq:Interaction Picture}
i\frac{d}{dt}\ket{\Phi(t)}=
\left(V^{\dagger}HV-iV^{\dagger}\frac{d}{dt}V\right)\ket{\Phi(t)}.
\end{equation}
This is in general called the interaction picture.

\vspace{3mm}
In the following we impose some restrictions on the model. 
Namely, we take $\omega(t)$ as in \cite{Dodonovs} 
\[
\omega(t)=\omega_{0}(1+\epsilon\sin(\eta t))
\]
where $\omega_{0}$, $\epsilon$ and $\eta$ are real constants. 
We assume that $\omega_{0}>0$, $0<\epsilon \ll 1$ and $\eta$ 
is large enough. 
Then $\omega(t)\approx \omega_{0}$ and
\[
\chi(t)
=\frac{\epsilon\eta\cos(\eta t)}{4(1+\epsilon\sin(\eta t))}
\approx \frac{\epsilon\eta}{4}\cos(\eta t)
\]
, and here we take $V$ as
\begin{equation}
\label{eq:V}
V=V(t)
=e^{-it\frac{\eta}{4}\sigma_{3}}\otimes e^{-it\frac{\eta}{2}N}
=\left(
\begin{array}{cc}
e^{-it\frac{\eta}{4}-it\frac{\eta}{2}N} &             \\
      & e^{it\frac{\eta}{4}-it\frac{\eta}{2}N}      
\end{array}
\right).
\end{equation}

Some calculation by use of (\ref{eq:V}) gives
\begin{eqnarray*}
\widehat{H}(t)&\equiv &V^{\dagger}HV-iV^{\dagger}\frac{d}{dt}V  \\
&=&
\left(\omega_{0}-\frac{\eta}{2}\right)1_{2}\otimes N+
i\frac{\epsilon\eta}{8}1_{2}\otimes 
\left\{(a^{\dagger})^{2}-a^{2}+e^{2i\eta t}(a^{\dagger})^{2}-e^{-2i\eta t}a^{2}\right\}  \\
&{}&+
\frac{\Omega-\eta/2}{2}\sigma_{3}\otimes {\bf 1}+
g(\sigma_{+}\otimes a+\sigma_{-}\otimes a^{\dagger})
\end{eqnarray*}
and we can also use the rotating wave approximation \ $e^{2i\eta t}\approx 0$\ \ 
when $\eta$ is large enough. As a result we finally obtain 
the time-independent Hamiltonian (see \cite{Dodonovs}) like
\begin{equation}
\label{eq:time-indep-Hamiltonian}
\widehat{H}=
\left(\omega_{0}-\frac{\eta}{2}\right)1_{2}\otimes N+
i\frac{\epsilon\eta}{8}1_{2}\otimes \left\{(a^{\dagger})^{2}-a^{2}\right\}
+
\frac{\Omega-\eta/2}{2}\sigma_{3}\otimes {\bf 1}+
g(\sigma_{+}\otimes a+\sigma_{-}\otimes a^{\dagger}).
\end{equation}

\vspace{5mm}\noindent
(v)\ Our Target.\ The aim of this paper is to solve 
the equation
\begin{equation}
\label{eq:Real Equation}
i\frac{d}{dt}\ket{\Phi(t)}=\widehat{H}\ket{\Phi(t)}
\end{equation}
under the general setting and the formal solution is given by
\begin{equation}
\label{eq:formal solution}
\ket{\Phi(t)}=e^{-it\widehat{H}}\ket{\Phi(0)}.
\end{equation}
Therefore, what we must do in the following is to calculate 
the term $e^{-it\widehat{H}}$ explicitly.

Before closing this section we must present an important 
problem.

\vspace{3mm}\noindent
{\bf Problem}\ \ We used approximation (a kind of rotating 
wave approximation) two times. 
Make an adaptive range (or region) of the model clear.

\section{Approximate Solution}
In this section we try to calculate the term $e^{-it\widehat{H}}$ 
in (\ref{eq:formal solution}). 
For simplicity we set
\[
-it\widehat{H}=X+Y
\]
where
\begin{eqnarray}
\label{eq:A and B}
&&X=-itA,\quad A=\left(\omega_{0}-\frac{\eta}{2}\right)1_{2}\otimes N+
i\frac{\epsilon\eta}{8}1_{2}\otimes \left\{(a^{\dagger})^{2}-a^{2}\right\}, 
\nonumber \\
&&Y=-itB,\quad B=\frac{\Omega-\eta/2}{2}\sigma_{3}\otimes {\bf 1}+
g(\sigma_{+}\otimes a+\sigma_{-}\otimes a^{\dagger}).  
\end{eqnarray}

For our purpose the Zassenhaus formula is useful :

\vspace{3mm}\noindent
Zassenhaus Formula\ \ We have an expansion
\[
\label{eq:Zassenhaus formula}
e^{X+Y}=
\cdots 
e^{-\frac{1}{6}\{2[[X,Y],Y]+[[X,Y],X]\}}
e^{\frac{1}{2}[X,Y]}
e^{Y}
e^{X}.
\]
Note that the formula is a bit different from that of \cite{CZ}. 
In this paper we use 
\begin{equation}
\label{eq:short Zassenhaus formula}
e^{X+Y}
\approx e^{\frac{1}{2}[X,Y]}e^{Y}e^{X}
=e^{-\frac{t^{2}}{2}[A,B]}e^{-itB}e^{-itA}.
\end{equation}
Therefore, let us calculate each term in the following.

\vspace{5mm}\noindent
[I]\ \ First, we calculate $e^{-itA}$. If we set 
\begin{equation}
\label{eq:alpha and beta}
\alpha=2\left(\omega_{0}-\frac{\eta}{2}\right),\ \ 
\beta=\frac{1}{4}\epsilon\eta
\end{equation}
then $A$ in (\ref{eq:A and B}) becomes
\[
A
=\frac{\alpha}{2}1_{2}\otimes N+
i\frac{\beta}{2}1_{2}\otimes \left\{(a^{\dagger})^{2}-a^{2}\right\}
=1_{2}\otimes\left\{
\alpha\frac{1}{2}N+
i\beta \left(\frac{1}{2}(a^{\dagger})^{2}-\frac{1}{2}a^{2}\right)\right\}.
\]

Here, we use a well--known Lie algebraic method, see for example \cite{KF}. 
Namely, if we set
\begin{equation}
\label{eq:su(1,1)-generators}
K_{+}=\frac{1}{2}(a^{\dagger})^{2},\quad
K_{-}=\frac{1}{2}a^{2},\quad
K_{3}=\frac{1}{2}\left(N+\frac{1}{2}\right)
\end{equation}
it is easy to see both $K_{+}^{\dagger}=K_{-},\ K_{3}^{\dagger}=K_{3}$ 
and the $su(1,1)$ relations
\begin{equation}
\label{eq:su(1,1)-relations}
[K_{3},K_{+}]=K_{+},\quad [K_{3},K_{-}]=-K_{-},\quad [K_{+},K_{-}]=-2K_{3}
\end{equation}
by use of the relation $[a,a^{\dagger}]={\bf 1}$. Then, $A$ in 
(\ref{eq:A and B}) can be written as
\begin{eqnarray*}
A
&=&1_{2}\otimes \left\{-\frac{1}{4}\alpha+
\alpha K_{3}+i\beta \left(K_{+}-K_{-}\right)\right\} \\
&=&
\left(
\begin{array}{cc}
-\frac{1}{4}\alpha+\alpha K_{3}+i\beta \left(K_{+}-K_{-}\right) &   \\
 & -\frac{1}{4}\alpha+\alpha K_{3}+i\beta \left(K_{+}-K_{-}\right)
\end{array}
\right)
\end{eqnarray*}
and from this we have
\[
e^{-itA}=
\left(
\begin{array}{cc}
e^{\frac{it}{4}\alpha}e^{-it\left\{\alpha K_{3}+i\beta \left(K_{+}-K_{-}\right)\right\}} &    \\
  & e^{\frac{it}{4}\alpha}e^{-it\left\{\alpha K_{3}+i\beta \left(K_{+}-K_{-}\right)\right\}}
\end{array}
\right).
\]
Therefore, we have only to calculate the term
\[
U(t)
\equiv e^{-it\left\{\alpha K_{3}+i\beta \left(K_{+}-K_{-}\right)\right\}}
= e^{-it\alpha K_{3}+t\beta \left(K_{+}-K_{-}\right)}.
\]

For the purpose we want to look for the following disentangling form
\begin{equation}
\label{eq:disentangling form}
U(t)=e^{f(t)K_{+}}e^{g(t)K_{3}}e^{h(t)K_{-}}
\end{equation}
with functions $f(t),\ g(t),\ h(t)$ ($f(0)=g(0)=h(0)=0$)\footnote
{this is a standard method}.

The result is as follows.

\begin{eqnarray}
\label{eq:f & g & h}
f(t)&=&
\frac{\frac{\beta}{\sqrt{-\frac{\alpha^{2}}{4}+\beta^{2}}}\sinh\left(t\sqrt{-\frac{\alpha^{2}}{4}+\beta^{2}}\right)}
{\cosh\left(t\sqrt{-\frac{\alpha^{2}}{4}+\beta^{2}}\right)+\frac{i\frac{\alpha}{2}}{\sqrt{-\frac{\alpha^{2}}{4}+\beta^{2}}}
\sinh\left(t\sqrt{-\frac{\alpha^{2}}{4}+\beta^{2}}\right)}, \nonumber \\
g(t)&=&
-2\log\left(
\cosh\left(t\sqrt{-\frac{\alpha^{2}}{4}+\beta^{2}}\right)+\frac{i\frac{\alpha}{2}}{\sqrt{-\frac{\alpha^{2}}{4}+\beta^{2}}}
\sinh\left(t\sqrt{-\frac{\alpha^{2}}{4}+\beta^{2}}\right) \right), \\
h(t)&=&
\frac{-\frac{\beta}{\sqrt{-\frac{\alpha^{2}}{4}+\beta^{2}}}\sinh\left(t\sqrt{-\frac{\alpha^{2}}{4}+\beta^{2}}\right)}
{\cosh\left(t\sqrt{-\frac{\alpha^{2}}{4}+\beta^{2}}\right)+\frac{i\frac{\alpha}{2}}{\sqrt{-\frac{\alpha^{2}}{4}+\beta^{2}}}
\sinh\left(t\sqrt{-\frac{\alpha^{2}}{4}+\beta^{2}}\right)}=-f(t). \nonumber
\end{eqnarray}
The derivation is analogous to that of \cite{Kazuyuki}. See also the appendix.

\vspace{5mm}\noindent
[II]\ \ Second, we calculate $e^{-itB}$.  The result is well--known 
and is given by
\begin{eqnarray}
\label{eq:calculation of B}
e^{-itB}
&=&
\exp
\left\{-it
\left(
\begin{array}{cc}
\frac{\Omega-\eta/2}{2} & ga                \\
ga^{\dagger} & -\frac{\Omega-\eta/2}{2}
\end{array}
\right)
\right\} \nonumber \\
&=&
\left(
\begin{array}{cc}
\cos t\sqrt{\varphi+g^{2}}-\frac{i\delta}{2}\frac{\sin t\sqrt{\varphi+g^{2}}}{\sqrt{\varphi+g^{2}}} & 
-ig\frac{\sin t\sqrt{\varphi+g^{2}}}{\sqrt{\varphi+g^{2}}}a \\
-ig\frac{\sin t\sqrt{\varphi}}{\sqrt{\varphi}}a^{\dagger} &
\cos t\sqrt{\varphi}+\frac{i\delta}{2}\frac{\sin t\sqrt{\varphi}}{\sqrt{\varphi}}
\end{array}
\right)
\end{eqnarray}
where we have set
\[
\delta\equiv \Omega-\frac{\eta}{2},\quad 
\varphi\equiv \frac{\delta^{2}}{4}+g^{2}N
\]
for simplicity.

\vspace{5mm}\noindent
[III]\ \ Third, we calculate $e^{-\frac{t^{2}}{2}[A,B]}$. 
From (\ref{eq:A and B}) simple calculation gives
\begin{eqnarray}
[A,B]
&=&
g\frac{\alpha}{2}(-\sigma_{+}\otimes a+\sigma_{-}\otimes a^{\dagger})-
ig\beta(\sigma_{+}\otimes a^{\dagger}+\sigma_{-}\otimes a)  \nonumber \\
&=&
g
\left(
\begin{array}{cc}
0 & -\frac{\alpha}{2}a-i\beta a^{\dagger} \\
\frac{\alpha}{2}a^{\dagger}-i\beta a & 0
\end{array}
\right)  \nonumber \\
&\equiv&
g
\left(
\begin{array}{cc}
0               & -D \\
D^{\dagger} & 0
\end{array}
\right) 
\end{eqnarray}
where we have set
\[
D=\frac{\alpha}{2}a+i\beta a^{\dagger}, \quad
D^{\dagger}=\frac{\alpha}{2}a^{\dagger}-i\beta a
\]
for simplicity. Note that
\[
[D,D^{\dagger}]=\left(\frac{\alpha^{2}}{4}-\beta^{2}\right){\bf 1}.
\]

The result is
\begin{equation}
e^{-\frac{t^{2}}{2}[A,B]}
=
\left(
\begin{array}{cc}
\cos\left(g\frac{t^{2}}{2}\sqrt{DD^{\dagger}}\right) & 
\frac{1}{\sqrt{DD^{\dagger}}}\sin\left(g\frac{t^{2}}{2}\sqrt{DD^{\dagger}}\right)D                   \\
-\frac{1}{\sqrt{D^{\dagger}D}}\sin\left(g\frac{t^{2}}{2}\sqrt{D^{\dagger}D}\right)D^{\dagger} & 
\cos\left(g\frac{t^{2}}{2}\sqrt{D^{\dagger}D}\right)
\end{array}
\right).
\end{equation}

\vspace{5mm}\noindent
[IV]\ \ As a result, our approximate solution to the 
equation (\ref{eq:Schrodinger}) is given by
\begin{equation}
\label{eq: approximate solution}
\ket{\Psi(t)}\approx e^{-\frac{t^{2}}{2}[A,B]}e^{-itB}e^{-itA}\ket{\Psi(0)}
\end{equation}
under any initial value $\ket{\Psi(0)}$. If we write
\begin{equation}
\label{eq:matrix form}
\widehat{U}(t)\equiv e^{-\frac{t^{2}}{2}[A,B]}e^{-itB}e^{-itA}
=
\left(
\begin{array}{cc}
U_{11} & U_{12} \\
U_{21} & U_{22}
\end{array}
\right)
\end{equation}
then each component is given by
\begin{eqnarray*}
&&U_{11}=e^{\frac{it}{4}\alpha}
\left\{
\cos\left(g\frac{t^{2}}{2}\sqrt{DD^{\dagger}}\right)
\left(\cos t\sqrt{\varphi+g^{2}}-\frac{i\delta}{2}\frac{\sin t\sqrt{\varphi+g^{2}}}{\sqrt{\varphi+g^{2}}}\right)-
\right.  \\
&&
\left.\qquad \quad\ \
ig\frac{1}{\sqrt{DD^{\dagger}}}\sin\left(g\frac{t^{2}}{2}\sqrt{DD^{\dagger}}\right)D\ 
\frac{\sin t\sqrt{\varphi}}{\sqrt{\varphi}}a^{\dagger}
\right\}e^{f(t)K_{+}}e^{g(t)K_{3}}e^{h(t)K_{-}}, \\
&&U_{12}=e^{\frac{it}{4}\alpha}
\left\{
-ig\cos\left(g\frac{t^{2}}{2}\sqrt{DD^{\dagger}}\right)
\frac{\sin t\sqrt{\varphi+g^{2}}}{\sqrt{\varphi+g^{2}}}a\ +
\right. \\
&&
\left.\qquad \quad\ \
\frac{1}{\sqrt{DD^{\dagger}}}\sin\left(g\frac{t^{2}}{2}\sqrt{DD^{\dagger}}\right)D
\left(\cos t\sqrt{\varphi}+\frac{i\delta}{2}\frac{\sin t\sqrt{\varphi}}{\sqrt{\varphi}}\right)
\right\}e^{f(t)K_{+}}e^{g(t)K_{3}}e^{h(t)K_{-}}, \\
&&U_{21}=e^{\frac{it}{4}\alpha}
\left\{
-\frac{1}{\sqrt{D^{\dagger}D}}\sin\left(g\frac{t^{2}}{2}\sqrt{D^{\dagger}D}\right)D^{\dagger}
\left(\cos t\sqrt{\varphi+g^{2}}-\frac{i\delta}{2}\frac{\sin t\sqrt{\varphi+g^{2}}}{\sqrt{\varphi+g^{2}}}\right)-
\right. \\
&&
\left.\qquad \quad\ \
ig\cos\left(g\frac{t^{2}}{2}\sqrt{D^{\dagger}D}\right)
\frac{\sin t\sqrt{\varphi}}{\sqrt{\varphi}}a^{\dagger}
\right\}e^{f(t)K_{+}}e^{g(t)K_{3}}e^{h(t)K_{-}}, \\
&&U_{22}=e^{\frac{it}{4}\alpha}
\left\{
ig\frac{1}{\sqrt{D^{\dagger}D}}\sin\left(g\frac{t^{2}}{2}\sqrt{D^{\dagger}D}\right)D^{\dagger}\ 
\frac{\sin t\sqrt{\varphi+g^{2}}}{\sqrt{\varphi+g^{2}}}a\ +
\right. \\
&&
\left.\qquad \quad\ \
\cos\left(g\frac{t^{2}}{2}\sqrt{D^{\dagger}D}\right)
\left(\cos t\sqrt{\varphi}+\frac{i\delta}{2}\frac{\sin t\sqrt{\varphi}}{\sqrt{\varphi}}\right)
\right\}e^{f(t)K_{+}}e^{g(t)K_{3}}e^{h(t)K_{-}}.
\end{eqnarray*}

\noindent
This is our main result.

\section{Closing Remarks}
In this paper we treated the model by A. V. Dodonov and V. V. Dodonov 
and presented another approach based on Mathematical Physics and 
obtained some analytic approximate solution. 
Details with applications and developments will be published separately.

We have neglected Dissipation (for example, the Cavity loss) in this 
paper, which is not realistic. If we take dissipation into consideration 
the model will become very complicated. For example, see our papers 
\cite{Kazuyuki}, \cite{EFS}, \cite{FS} and \cite{FS-1}, \cite{FS-2} 
(the last two are highly recommended). 

Then, to obtain an analytic approximate solution will become increasingly 
difficult. Further study and new ideas are needed.

\vspace{10mm}
\begin{center}
 \begin{Large}
  {\bf Appendix}
 \end{Large}
\end{center}
\vspace{5mm}
A note is added. \ In the text we must calculate
\[
U(t)
=e^{-it\left\{\alpha K_{3}+i\beta \left(K_{+}-K_{-}\right)\right\}}
=e^{-it\alpha K_{3}+t\beta \left(K_{+}-K_{-}\right)}.
\]
From this we have the differential equation
\[
\frac{d}{dt}U(t)=\left\{-i\alpha K_{3}+\beta (K_{+}-K_{-})\right\}U(t).
\]
On the other hand, under the ansatz (\ref{eq:disentangling form}) 
(: $U(t)=e^{f(t)K_{+}}e^{g(t)K_{3}}e^{h(t)K_{-}}$) some calculation gives
\[
\frac{d}{dt}U(t)=
\left\{
\left(\dot{f}-\dot{g}f+\dot{h}e^{-g}f^{2}\right)K_{+}+
\left(\dot{g}-2\dot{h}e^{-g}f\right)K_{3}+
\left(\dot{h}e^{-g}\right)K_{-}
\right\}U(t)
\]
where we have used $\dot{f}=\frac{df}{dt}$, etc for simplicity. 
Therefore, by comparing two equations we obtain
\[
\left\{
\begin{array}{l}
\dot{f}-\dot{g}f+\dot{h}e^{-g}f^{2}=\beta \\
\dot{g}-2\dot{h}e^{-g}f=-i\alpha \\
\dot{h}e^{-g}=-\beta
\end{array}
\right.
\Longrightarrow\ 
\left\{
\begin{array}{l}
\dot{f}-\dot{g}f-\beta f^{2}=\beta \\
\dot{g}+2\beta f=-i\alpha \\
\dot{h}e^{-g}=-\beta
\end{array}
\right.
\Longrightarrow\ 
\left\{
\begin{array}{l}
\dot{f}+i\alpha f+\beta f^{2}=\beta \\
\dot{g}+2\beta f=-i\alpha \\
\dot{h}=-\beta e^{g}.
\end{array}
\right.
\]
The equation
\[
\dot{f}+i\alpha f+\beta f^{2}=\beta
\]
is a (famous) Riccati equation of general type. 
If we can solve the equation we have the solutions like \ 
$f(t)\Longrightarrow g(t)\Longrightarrow h(t)$. 
See (\ref{eq:f & g & h}) as these solutions.

\vspace{10mm}
%%%%%%%%%%%%%
%References%
%%%%%%%%%%%%%

\end{document}